\documentclass[%
 amsmath,amssymb,
 aps,nofootinbib,
 reprint,%
]{revtex4-2}

\usepackage{graphicx}
\usepackage{dcolumn}
\usepackage{bm}

\usepackage[utf8]{inputenc}
\usepackage[T1]{fontenc}
\usepackage{mathptmx}
\usepackage{etoolbox}
\usepackage{xcolor}
\usepackage{float}

\makeatletter
\def\@email#1#2{%
 \endgroup
 \patchcmd{\titleblock@produce}
  {\frontmatter@RRAPformat}
  {\frontmatter@RRAPformat{\produce@RRAP{*#1\href{mailto:#2}{#2}}}\frontmatter@RRAPformat}
  {}{}
}%
\makeatother
\begin{document}

\preprint{APS/123-QED}

\title{An N-loop potential energy model for levitated mm-scale magnets in cm-scale superconducting coaxial microwave resonators}

\affiliation{Physics, School of Natural Sciences, University of California, Merced, 95343 CA, USA.}
\affiliation{Lawrence Livermore National Laboratory, Livermore, 94550, CA, USA.}

\author{Jeffrey Miller$^1$}

\author{Nabin K. Raut$^1$}%

\author{Demitrius Zulevic$^1$}

\author{Harold Hart$^1$}

\author{Luis A. Martinez$^2$}

\author{Alessandro Castelli$^2$}

\author{Raymond Chiao$^1$}

\author{Jay E. Sharping$^1$}

\email{jsharping@ucmerced.edu}

\date{\today}

\begin{abstract}
The levitation of a macroscopic object within a superconducting resonator provides a unique and novel platform to study optomechanics, quantum information, and gravitational wave detection. Existing mirror-method and single-loop models for calculating magnet levitation are insufficient for predicting the position and motion of the levitated magnet. If the cavity-magnet interaction is modeled using a large number of smaller surface current loops, one can quantitatively model the dynamics of the levitation of the magnet within the cavity. The magnet's most-likely position and orientation can be predicted for non-trivial cavity geometries and cavity orientations. Knowing the potential energy landscape within the cavity configuration also provides a means to estimate the resonant mechanical frequencies at which the levitated magnet vibrates, and enables tailoring the cavity design for specific outcomes.
\end{abstract}

\maketitle
 \section{Introduction}
Superconducting cavities are a key element in the current generation of quantum computers~\cite{Reagor2016}. Cavities coupled to transmon qubits allow the readout of quantum states via the change in cavity resonance frequency~\cite{Blais07, Schmitt14}. In the cm-scale, superconducting coaxial stub cavities provide a high-Q (fundamental) resonant electromagnetic mode, the quarter-wave resonance. This quarter-wave resonance is easily isolated from harmonics or other modes in the 1-10~GHz frequency range, which makes them ideal for qubit spectroscopy. The cavity mode couples in such a way as to enhance the lifetime and allow for state readout of transmon qubits~\cite{Reagor2016, Kutsaev2020}. Stub cavities are also useful in coupling to the quantum spectrum of other objects, such as magnons~\cite{Goryachev2014,Zhang2014}. The spectroscopic properties of high-Q cavities are broadly useful beyond quantum computing, however. The position and width of the narrow spectral line is sensitive to any perturbation of the electromagnetic field within the cavity. Thus, bulk cavities are used in efforts to detect dark matter, search for axions and detect gravitational waves~\cite{McAllister_2016_b,McAllister_2017,McAllister_2017_b,McAllister_2018,McAllister_2019,GORYACHEV2018,Tobar_2022,Quiskamp_2020,Quiskamp2022}.

Quantum magnetomechanics is a field of study that explores the interaction between quantum mechanical systems and mechanical motion in the presence of magnetic fields~\cite{Cirio2012quantum}, Frequently one uses magnetic forces to trap or levitate a particle and damp its motion so that it approaches its ground state. This interdisciplinary field combines concepts from quantum physics, magnetism, and mechanical engineering to explore phenomena such as quantum state manipulation and quantum-coherent mechanical oscillations. The fundamental ideas in magnetomechanics are the same as those presented for optomechanical systems: trapping or suspending particles, understanding the thermodynamical behavior associated with those particles, using active feedback or passive methods to cool the center-of-mass motion, and applying the system to force sensing~\cite{Millen2020optomechanics}. In the quantum realm, one envisions potential tests of quantum physics using massive objects and coupling to other quantized degrees of freedom, such as spins in nitrogen-vacancy centers, via the levitated micro- or nano-particles~\cite{Gieseler2020single-spin,Pan2023enhanced}. A recent review of the mechanics of magnet levitation above type I and type II superconductors by Vinante, et al., summarizes the current state-of-the-art and various potential applications~\cite{Vinante2022levitated}. 

The work of Giesler, \textit{et al.}~\cite{Gieseler2020single-spin}. Pan, \textit{et al.}~\cite{Pan2023enhanced}. and Prat-Cramps, \textit{et al.}~\cite{Prat-camps2017ultrasensitive}. are good examples of real and proposed systems employing levitated magnets. Here we investigate the combination of levitated magnets and superconducting cavities with a vision of taking advantage of electromagnetic field localization and spectral sensitivity as a means for precision metrology. We report on a potential energy model for a system in which a millimeter-scale permanent magnet is levitated inside of a superconducting coaxial microwave ($\sim10$~GHz) resonator. Knowing the potential energy of the system allows one to relate measured frequency shifts as a function of temperature to possible phenomena that occur inside of a cavity that has no optical access. We have used this model to complete several numerical studies that answer the following questions in section 3:

    \begin{enumerate}
        \item What is the most likely position of stable levitation inside of the original cavity design?
        \item Can we design a cavity such that positions of global minima can be controlled?
        \item Which magnet orientation yields the lowest potential energy?
    \end{enumerate} 
    
The N-loop model detailed below provides a way to determine the most likely orientation and levitation height of a magnet in a given cavity configuration, including configurations outside of the coaxial case presented here. We show that the model can aid us in designing cavities in a way that allows us to force the magnet to levitate in specific regions of the cavity to improve frequency sensitivity to motion. Using the model also allows us to estimate the mechanical frequency at which the magnet vibrates along each axis while levitating by modeling the potential well as a spring potential.

Superconducting microwave resonators are metal structures that confine electromagnetic radiation at microwave frequencies. On resonance within the structure, the waves interfere constructively and form stationary waves inside of the cavity. These types of cavities are important for applications such as particle accelerators due to the extremely low loss exhibited in superconducting resonant structures~\cite{Padamsee1992}. Another attractive feature of a bulk RF cavity is the extremely narrow bandwidth that allows it to work as a highly selective band-pass filter~\cite{Nguyen2007}. The quarter wave cavity design can be specifically used to improve qubit lifetime by placing a qubit based on a Josephson junction inside of a high Q quarter wave resonator~\cite{Kutsaev2020,Aspelmeyer2014}. Mechanical force sensitivity on the order of $3 \frac{aN}{\sqrt{Hz}}$ has been achieved by placing a nano-mechanical beam inside of a superconducting microwave cavity~\cite{Regal2008}.

The coaxial design used in levitation experiments (data compared with our model in section D) is modeled as a transmission line that is open on one end and grounded at the opposite end. Fig. 1a shows the coaxial design used in these experiments. The body of the cavity has an outer radius of 7 mm and a height of 55 mm, and the coaxial stub is connected to the bottom surface (the short) with a height of 5 mm and a radius of 2 mm. Unlike many SRF resonator designs, the exponential decay of the electric field near the coaxial stub allows for an open end and removes much of the loss present near the seams that closed designs possess. The physical configuration of the mode of interest is important in that it leads to a resonant frequency that is highly sensitive to small perturbations near the stub. The increased sensitivity makes the motion of the levitated particle visible in the form of a frequency shift away from the bare cavity resonance. This cavity design is sometimes called a quarter-wave stub cavity because the resonance frequency can be roughly approximated as $f_0 = \frac{c}{4L}$, where $L$ is the length of the coaxial stub and $c$ is the speed of light~\cite{Sharping2020}. In practice, the physical dimensions of the stub height, stub radius, and cavity radius determine the measured resonance with an inverse proportionality. As the stub size (both radius and height) decreases, the resonance frequency increases and vice versa. Multiple stub parameters have been fabricated and tested with each of them resulting in a measured $f_0$ between $9-11$ GHz. These cavity parameters defined one of the main goals in developing this model, which is to design a cavity with dimensions such that the magnet has a minimum potential energy near top of the stub to allow more sensitive frequency shifts to characterize motion where there is no optical access inside of the cavity.

\begin{figure}[ht]
\includegraphics[scale=1]{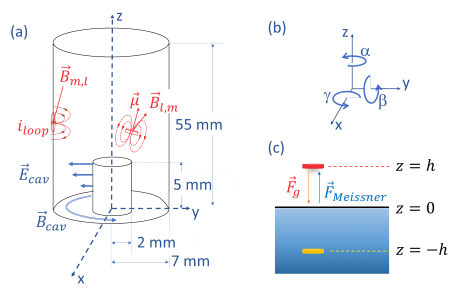}
\caption{\label{fig:Model}(a) Schematic of the coaxial quarter-wave stub cavity showing the cavity geometry, electric and magnetic fields of the cavity mode, and the model used to evaluate the magnet and loop magnetic fields. (b) The definition of the rotation angles used in the magnet tilt calculations where everything is defined from the center of mass of the magnet. (c) Illustration of the image method wherein a virtual magnet is situated symmetrically below an infinite plane such that the magnets repel one another.}
\end{figure}

\begin{table}
\caption{\label{keyInfo}Cavity and disk parameters}
\begin{ruledtabular}
\begin{tabular}{lcr}
Quantity&Value\\
\hline
        RF mode resonance & 10~GHz\\
        Cavity height & 55~mm\\
        Cavity radii & 5~mm--7~mm\\
        Stub height & 5~mm\\
        Stub radius & 2~mm\\
        Magnet radius & Disk=0.5~mm; Sphere=0.5~mm\\
        Disk magnet thickness & 0.5~mm\\
        Disk magnetization direction & axial\\
        Magnet mass & Disk=2.75~mg; Sphere=3.67~mg\\
        Magnet strengths & 1.22~T, 1.35~T, 1.44~T, 1.47~T\\
\end{tabular}
\end{ruledtabular}
\end{table}

By introducing a magnet into the coaxial stub cavity, finite element simulations can look at the effect it has on the resonance frequency at various configurations including height, radial distance from the center of the cavity, and the tilt of the magnet. These simulations are compared to experimental data recorded with the system inside a dilution refrigerator at $mK$ temperatures where direct, optical access to observe levitation is not possible in the dilution refrigerator configuration used for these measurements~\cite{Raut2021}. The potential energy landscape inside of the cavity geometry gives indirect evidence of where stable levitation occurs. One can calculate a rough estimate of these minimum energy points with the method of images or other simple models~\cite{Perez-Diaz2007,Arkadiev1947}, but the unique and non-trivial geometry of the cavity necessitates a stronger, more robust model that can find the minima for this particular configuration along with other generalized, non-trivial geometries. A comparison between these models is shown in Fig. 3 to look at levitation heights versus measured levitation height.

The magnets used in these simulations, as well as in the experimental data, are neodymium disk-shaped and spherical magnets with sub-mm dimensions (see Table~\ref{keyInfo}). They have remanent field strengths of 1.22 T, 1.35 T, 1.44 T, 1.47 T (N35, N42, N50, and N52). We identify the trend of increasing levitation height with increasing remanent field strengths in both experimental data and simulations. The following section outlines a comparison of models of increasing complexity, starting from a method of images and finishing with a N-loop model that takes the entire geometry of the superconducting cavity into account.

\section{Levitation Modeling}
\subsection{Image Method}

    The method of images used in many problems in electromagnetism can be adapted to estimate the stable levitation height above a superconducting plane~\cite{Lugo1999}. In the simple system shown in Fig.~\ref{fig:Model}(c), stable levitation occurs when the downward force due to gravity is equal to the levitation force caused by the Meissner effect. Setting these two forces equal allows for the extraction of the levitation height.
    
    \begin{equation}
        |\vec{F}_{Meissner}| = \frac{3\mu_0|\vec{\mu}|^2}{32\pi z^4} = |\vec{F}_g| = \rho V g,
    \end{equation}
    \begin{equation}
        |\vec{\mu}|^2 = \left(\frac{B_r V}{\mu_0}\right)^2,
    \end{equation}
\noindent where, $\rho$ is the density of the material of the magnet, $V$ is the volume of the magnet, and $B_r$ is the remanent field strength of the magnet.     
    
Solving Eqs.~(1) and (2) for $z$, we arrive at Eq.~(3) which is a rough estimate of the levitation height given the magnet parameters. 
    
    \begin{equation}
        z=\left(\frac{3V}{32\pi \mu_0 \rho g}\right)^{(1/4)} \sqrt{B_r}
    \end{equation}

The heights given by this model are plotted as a function of remanent field strength in Fig.~\ref{fig:ModelComparisons} and are compared with more advanced models. Using this simple model gives a rough estimate of how high the magnet is expected to levitate, however, the obvious limitation is that it is limited to extremely simple geometry, such as standard cylindrical cavities. Compared with experimental data~\cite{Raut2021}, the levitation height calculated with the image model is more than two times larger than the actual value.

\subsection{Two-loop model}
\begin{figure}[ht]
\includegraphics[scale=.25]{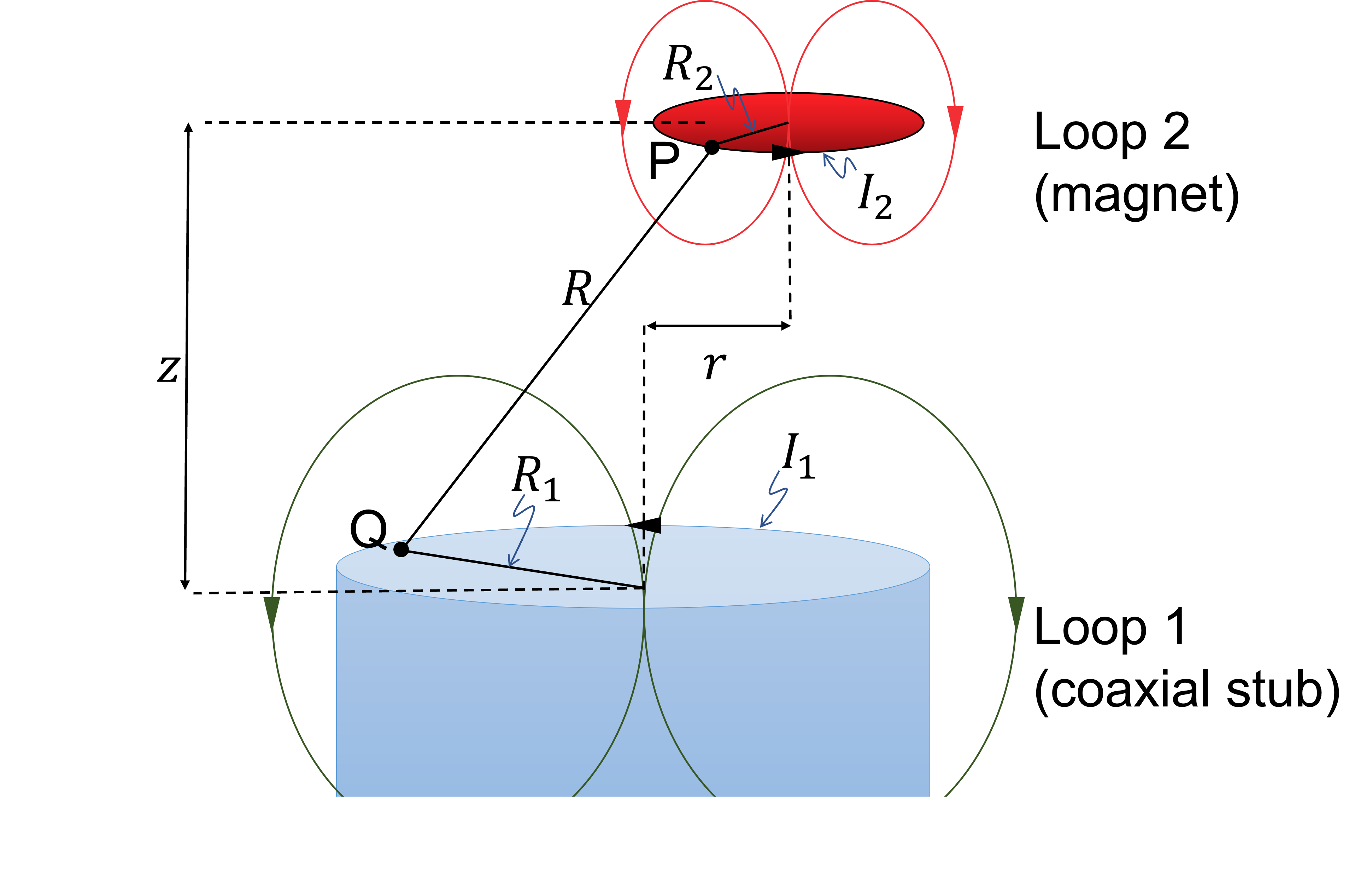}
\caption{\label{fig:2loopModel}The two-loop model. The coaxial stub (blue) is considered a current-carrying loop defined by the circumference of the stub, while the magnet is modeled as a solenoid. The approximate magnetic field lines are shown in red and blue.}
\end{figure}

Another model that can be used to predict the levitation height is a current loop-based model~\cite{Kim1996}. Figure~\ref{fig:2loopModel} shows the configuration for the current loop model where loop 1 represents the coaxial stub and loop 2 represents the magnet. Treating the cylindrical magnet as a solenoid~\cite{Cebron2021download}, we calculate the normal component of the magnetic field at the center of the stub ($B_\textrm{z,loop}$) as a function of the relative position of the magnet. The diamagnetic response of the stub is then the magnetic field of the current-carrying loop 1 of radius, $R_1$. The loop current is given by $I_{loop}=-B_\textrm{z,loop} R_1 / \mu_0$. The magnetic field at the first loop due to the second loop is calculated with the complete elliptic integrals of the first kind and second kind,

    \begin{equation}
        E = \int_0^{\frac{\pi}{2}} (1-\kappa^2sin^2\theta)^{\frac{1}{2}}d\theta
    \end{equation}
    \begin{equation}
        K = \int_0^{\frac{\pi}{2}} \frac{1}{(1-\kappa^2sin^2\theta)^{\frac{1}{2}}}d\theta,
    \end{equation}

\noindent with

    \begin{equation}
        B_r = \frac{\mu_0 I_1 z}{2\pi r[(R_1+r)^2+z^2]^{1/2}} \left[ \frac{R_1^2 + r^2 + z^2}{(R_1-r)^2+z^2}E-K\right]
    \end{equation}
    \begin{equation}
        B_{\phi} = 0
    \end{equation}
    \begin{equation}
        B_z = \frac{\mu_0 I_1}{2\pi r[(R_1+r)^2+z^2]^{1/2}} \left[ \frac{R_1^2 - r^2 - z^2}{(R_1-r)^2+z^2}E+K\right],
    \end{equation}

\noindent and the elliptic integral parameter $\kappa$ is given by $\kappa^2 = \frac{4R_1 r}{(R_1+r)^2+z^2}$, $r$ is the lateral offset between each loop, $I_\textrm{loop}$ is the current in the loop 1, and $z$ is the height of the magnet above the stub. The dot product of the vector magnetic field and the magnetic moment of the cylindrical magnet (now assumed to be a dipole) is taken to yield the potential energy,

    \begin{equation}
        U = -\frac{1}{2}\vec{\mu}\cdot\vec{B} + mgz.
        \label{Eq.2CoilPotential}
    \end{equation}

\noindent In Eq.~\ref{Eq.2CoilPotential}, $mgz$ is the gravitational potential energy of the magnet. The two-loop model predicts levitation heights which are similar to the method of images (see Fig.~\ref{fig:ModelComparisons}), but it ignores the walls and gap region of the cavity. 

\subsection{N-loop model}

To capture more of the cavity's geometric features, and to determine the stable equilibrium conditions for a magnet, we generalize the two-loop model into an 'N-loop' model. The superconducting surface is broken up into equally sized mesh elements. For each of the mesh elements, the magnetic field from the permanent magnet, $\vec{B}_{m,\ell}$, induces a small loop current, $i_\textrm{loop}$, as shown in Fig.~\ref{fig:Model}(a). The loop produces a magnetic field in response to the induced surface current, $\vec{B}_{\ell,m}$. This response field then acts back on the magnet. The field calculation from the two-loop model is used to calculate the response field arising from each of the loops. The contributions from each loop element are summed and the potential energy contribution of that particular loop current is again calculated according to Eq.~\ref{Eq.2CoilPotential}.

\begin{figure}[ht]
\includegraphics[scale=0.9]{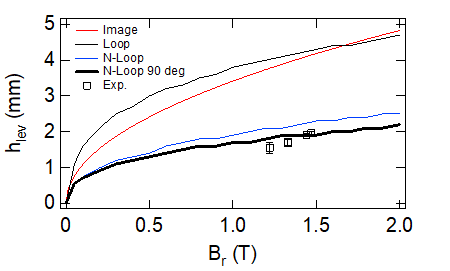}
\caption{\label{fig:ModelComparisons}Comparison between experimentally measured values and of the levitation heights predicted by the various. The N-loop models provide a more accurate prediction.}
\end{figure}

    Expanding the two-loop model into the N-loop model allows one to consider the non-trivial geometry of the cavity. Not only will this improved model numerically calculate the levitation height of the magnet (as shown in the comparisons with experimental measurements in Fig.~\ref{fig:ModelComparisons}), it is also used to generate a contour map of the potential energy with respect to the geometry of the cavity. This is crucial for fully characterizing the system as one is able to calculate the position of minimum potential energy to find the point for stable levitation inside of the cavity. This calculation can be used for arbitrary geometries, and to design cavity configurations that force levitation in regions of interest, namely the region above the coaxial stub that is most sensitive to small perturbations. 

\section{Results}
Five studies are detailed in the following section. In section III A we fix the magnet's position and confirm that the model produces the expected variation in potential energy as a function of magnetic orientation (\emph{i.e.,} direction of the magnetic moment). In Section III B, we present and evaluate the potential energy landscapes as a function of magnet position within the cavity. In Section III C, we return to a more quantitative investigation of magnet orientation in the potential well. In Section III D, we discuss how we use finite element calculation results to obtain levitation heights from experimentally-measured RF spectra. In Section III E, we investigate tilting the cavity to obtain a single minimum in the potential well. We then estimate the center-of-mass vibrational frequencies for a magnet levitated in such a tilted cavity.
\subsection{Validating the Model}
To test the N-loop model we verified its agreement with experimental data for magnet levitation for simple magnet-cavity configurations. One such configuration is a magnet placed at various heights above the coaxial stub along the cavity's cylindrical axis. These calculations include contributions from all interior surfaces of the cavity. The model accurately predicts levitation height compared with earlier measurements from our group~\cite{Raut2021}, as shown in Fig.~\ref{fig:ModelComparisons}. The two lower traces in Fig.~\ref{fig:ModelComparisons} are for the N-loop model for two different orientations of the magnetic moment. Both provide relatively good agreement with the data compared with the simpler models. The N-loop model predicts the experimental levitation heights within 5\%, while the image and single-loop models are off by nearly a factor of 2. 

\begin{figure}[ht]
\includegraphics[scale=1]{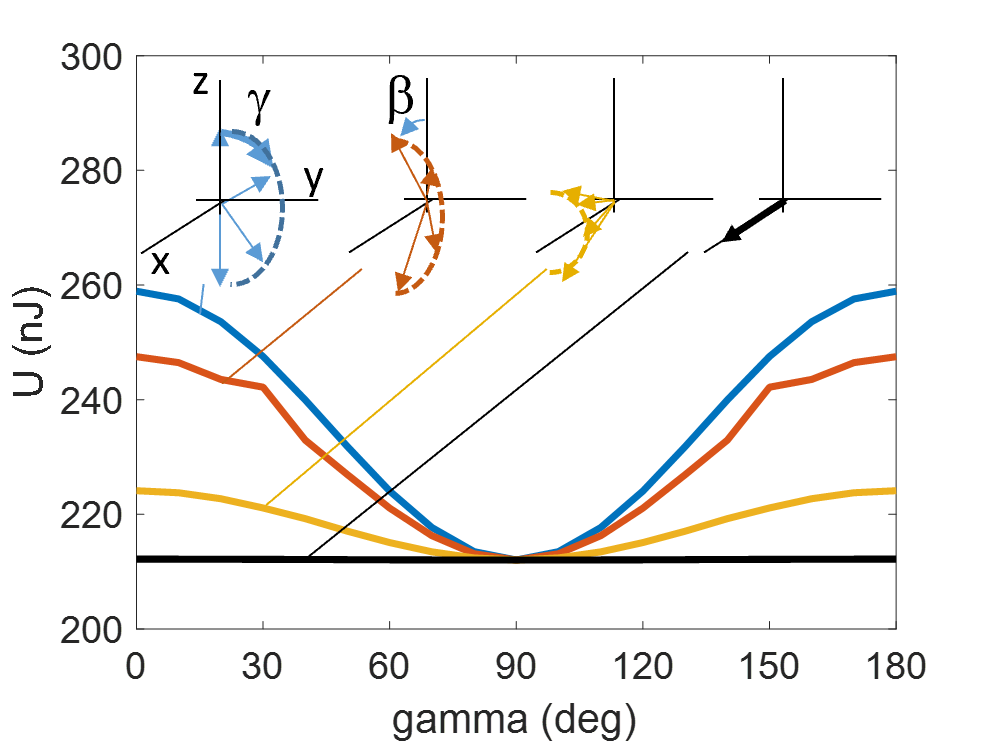}
\caption{\label{fig:validation}Plots showing the total potential energy as a function of moment angle for the cylindrical N52 magnet. Each trace is $U$ as a function of $\gamma$ for a different value of $\beta$ (rotation about the y-axis). The blue curve represents $\beta=0~^\textrm{o}$, while the black curve is $\beta=90~^\textrm{o}$}.
\end{figure}

The N-loop traces in Fig.~\ref{fig:ModelComparisons} also confirm that the predicted levitation heights differ slightly depending on the alignment of the magnetic moment. It is critical, then, to know the orientation of the magnetic moment within the cavity. We expect $\vec{\mu}$ to align horizontally, parallel with the closest surface~\cite{Lugo1999}. This is because the magnetic field decreases more quickly with distance in the direction perpendicular to the magnetic moment. Figure \ref{fig:validation} shows the total (magnetic and gravitational) potential energy as a function of orientation about the x-axis ($\gamma$) with the magnet located along the axis of the cavity, $2$ mm above the surface of the stub. Because the magnet's position is on the axis of the cavity, contributions from the vertical surfaces (outer wall and stub wall) and the floor of the cavity are small compared to contributions from the surface of the stub. Each trace corresponds to a particular value of $\beta$ as depicted in the inset sketches. The plots in Fig. \ref{fig:validation} show a minimum when $\gamma=90$, except for $\beta=0$. We see the expected potential energy response for a magnet positioned above a superconducting plane with a maximum when $\vec{\mu}$ is aligned with the positive z-axis (normal to the surface) and minima when $\vec{\mu}$ is parallel with the surface.

Having validated the N-loop model in comparison with our experimental results with cylindrical magnets, we now shift our attention to spherical magnets. Spherical magnets have the advantages of having a fully symmetric mass distribution as well as a dipole-shaped magnetic field distribution.

\subsection{Potential energy landscape}
In this section, we investigate the potential energy landscape for a spherically-shaped magnet to predict where it will levitate within the cavity and in what orientation. The switch to a spherical magnet allows the use of a simpler calculation of the magnetic field from the magnet at a surface element on the cavity surface. A spherical magnet allows for a dipole-like magnetic field calculation as opposed to modeling the cylindrical magnet as a solenoid.  We performed calculations for commonly-available magnet strengths, cavity sizes, magnet orientations and magnet positions, with the stub radius and height fixed at $r_{stub}=2$mm and $h_{stub}=5$mm. We evaluate the total potential energy over an array of magnet positions within the y-z plane.

\begin{figure}
\includegraphics[scale=.5]{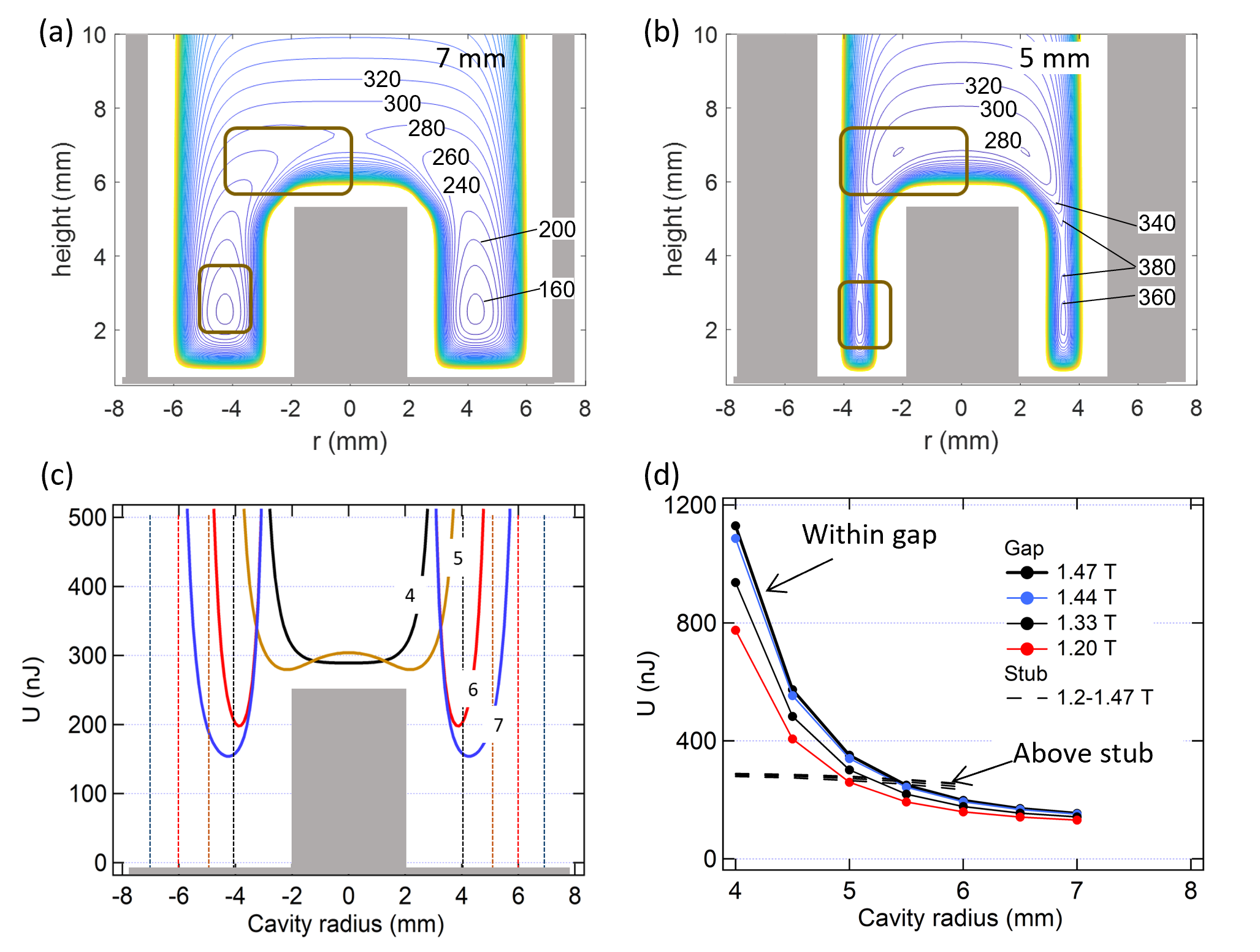}
\caption{\label{fig:study1}Levitation of a 1.47~T, spherically-shaped magnet whose magnetic field is directed azimuthally around the stub ($\gamma=0$ and $\beta=90^{\circ}$). (a) Contour plot showing the total potential energy as functions of cavity radial position and height for a magnet within a 7-mm-radius cavity. The global minimum is located in the gap between the stub and the outer wall. (b) Contour plots for the same magnet in a 5-mm-radius cavity reveal minima within the gap and off-center above the stub. The contour lines in (a) and (b) are spaced by 20 nJ and boxes highlight the regions of interest within the cavity. (c) The potential energy as a function of radial position  illustrates the migration of the global minimum from above the stub to within the gap as the cavity outer radius ranges from 4~mm-7~mm. (d) Potential energy for the gap minimum and above-stub minimum as a functions of outer cavity radius. For the (1.47~T) magnet, the global minimum is located above the stub for cavity radii less than $\sim5.5$~mm, and within the gap for radii greater than 5.5~mm.}
\end{figure}

Typical results for the N52 magnet with its magnetic moment aligned with $\gamma=0^\textrm{o}$, and $\beta=90^\textrm{o}$ are shown in Fig.~\ref{fig:study1}. The global minimum in the potential energy occurs when the magnetic moment is directed azimuthally around the stub ($\gamma=0^\textrm{o}$, and $\beta=90^\textrm{o}$). 
Figure~\ref{fig:study1}(a) is a  contour plot of the total potential energy for the vertically aligned N52 (1.47~T) magnet in a cavity having a radius of 7~mm. Figure~\ref{fig:study1}(b) is the same magnet in a 4.5-mm-radius cavity. The boxes in Figs.~\ref{fig:study1}(a) and (b) highlight two regions of interest in the potential energy, one within the gap and one near the corner of the stub. For this configuration, the potential well is symmetric about the cavity axis, so global and local minimums are also located symmetrically on the positive side of the cavity. For cavity radii less than 5.5~mm, a potential well forms above the edge of the stub, as shown in Fig.~\ref{fig:study1}(b). As the outer cavity radius is reduced from 7~mm down to 4~mm the potential energy at the bottom of the well located near $r=\pm4$~mm (within the gap) rises until it is larger than the value near $r=\pm2$~mm (above the stub) as shown in Fig.~\ref{fig:study1}(c) and Fig.~\ref{fig:study1}(d). As the cavity shrinks in radius, the global minimum in potential energy shifts from within the gap between the stub and the cavity wall to the region near the rim of the stub. Note also that the potential energy on the axis of the cavity forms a saddle point and thus an unstable equilibrium for levitation.

Figure~\ref{fig:study1}(c) shows the calculated potential energy as a function of the magnet's radial position for a vertically-aligned magnet at its steady-state levitation height. The 4 curves represent different cavity radii ranging from 4~mm up to 7~mm. The gray rectangle represents the stub. One can see how the potential well rises and shifts as the cavity radius gets smaller. We also show the minimum potential energy at the two locations (in the gap and above the stub) as a function of cavity radius in Fig.~\ref{fig:study1}(d). When the cavity radius is smaller than about 5.5~mm, the curves cross one another and the global minimum moves above the stub. For large-gap cavities, the global minimum is within the gap. Similar plots were generated with the three other magnets (N35, N42, N52). These data are plotted in Fig.~\ref{fig:study1}(d). The above-stub values for the minimum potential energy vary by only $\sim$20~nJ, which is roughly the thickness of the line.
    
The important conclusion drawn from this set of calculations is that one can locate a global minimum near the corner of the stub. This is a region of high electric-field concentration, and thus there is a large frequency sensitivity to small perturbations in the magnet's position.  

\subsection{Magnet Orientation}
\begin{figure}
\includegraphics[scale=.7]{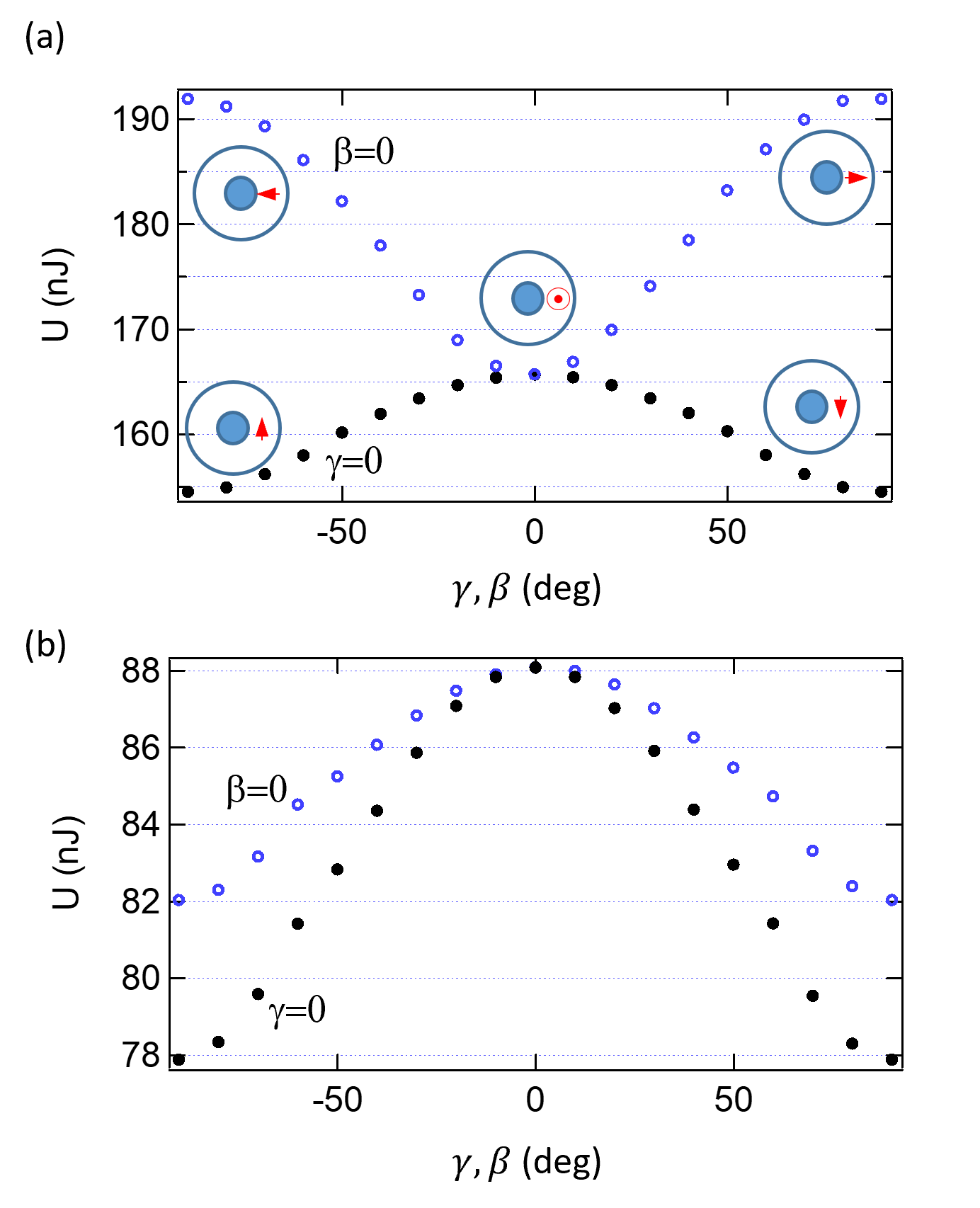}
\caption{\label{fig:UvsMagTilt}(a) Plots of minimum potential energy versus angle of the magnetic moment for a 1.47-T spherical magnet within a 7-mm radius stub cavity. Open circles ($\beta=0$) and $\gamma=-90^{\circ}-90^{\circ}$ is the independent variable on the x-axis. Filled circles ($\gamma=0$) and $\beta=-90^{\circ}-90^{\circ}$ is the independent variable on the x-axis. Insets show a top view of the direction of the magnetic moment for several configurations.(b) The same plots for a much weaker magnet, $B_r$=0.2~T.}
\end{figure}

The results of the evaluation of the preferred magnetic moment direction are presented in Fig.~\ref{fig:UvsMagTilt}. Figure~\ref{fig:UvsMagTilt}(a) shows the behavior of a 1.47-T spherical magnet as the angular direction of the magnet is varied. Note that to explore all orientations, it is sufficient to rotate the magnet about the x-axis, varying $\gamma$ while $\beta$ is fixed, and about the y-axis, varying $\beta$ while $\gamma$ is fixed. Inset sketches depict the magnet orientations, viewed from the top, for angles of 90$^{\circ}$ and 0$^{\circ}$ where the arrow indicates the direction of the magnetic moment. Each data point represents the global minima for the magnet in the cavity at the angle specified. The study is similar to that given in Fig.~\ref{fig:validation}, except that the magnet is now in the bottom of the deepest potential well, which is located in the gap between the stub and the cavity wall. 

The potential energy is found to be relatively large when the magnetic moment points directly towards the stub or the wall. When the magnetic moment is parallel to the axis of the cavity we find that, for a cavity with a radius of 7~mm, the vertically-oriented magnet experiences a saddle point in potential energy. The global minimum in the potential energy occurs when the magnetic moment is directed azimuthally around the stub (the bottom-left or bottom-right of the plot). Our model suggests, then, that a spherically-shaped, 1.47~T magnet levitating within a 7~mm cavity will come to rest in the gap of the cavity with its moment directed azimuthally around the stub. 

Interestingly, variation of U as a function magnet direction depends on the magnetic field strength. Figure~\ref{fig:UvsMagTilt}(b) shows the same plots for a much weaker magnet, $B_r=0.2$~T. In this case, while the configuration for the global minimum is the same, there is no longer a saddle point when $\gamma=\beta=0$. Two major observations from this set of calculations are that: 1) the magnet prefers to be directed azimuthally around the stub, and 2) the tip of the magnetic moment vector is pushed away from the nearest boundaries and favors pointing toward the most distant boundary. Furthermore, as shown in Fig.~\ref{fig:ModelComparisons}, the N-loop model provides a better prediction of the levitation heights we see experimentally. 

\subsection{Comparison with finite element Simulations and Measurements}
\begin{figure}[ht]
\includegraphics[scale=.8]{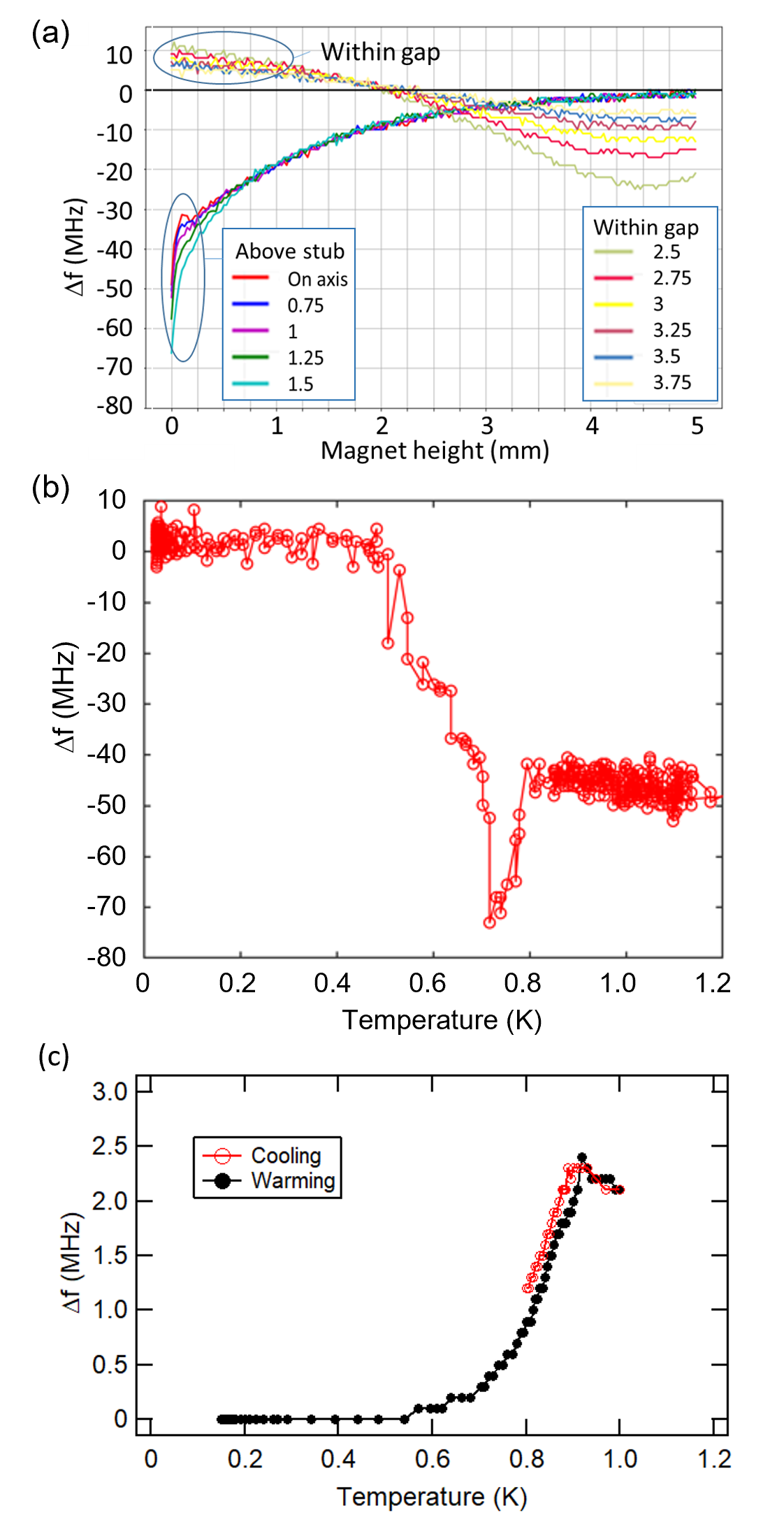}
\caption{\label{fig:HeightVsFrequency}a) Finite element simulation results describing the shift in resonance frequency when a disk magnet is levitated from the top of the stub or within the gap cavity at some radial offset from the center. b) Experimental measurements for a disk magnet within a 5-mm diameter stub cavity at low temperature. c) Experimental results for a spherical N52 magnet levitated above the floor of a 7-mm radius cavity. }
\end{figure}

Detailed discussions of several magnet levitation experiments are given in our previous work~\cite{Raut2021}. Here we provide an example of the use of the finite element simulations to interpret a particular set of data, as shown in Fig.~\ref{fig:HeightVsFrequency}. Recall that the resonant frequency of the $\lambda/4$ mode of the coaxial stub cavity is shifted by the presence of the magnet. Figure~\ref{fig:HeightVsFrequency}~(a) shows the results of finite element calculations of the shift in resonance frequency as a function of levitation height for a disk magnet at various radial positions within the 5-mm diameter cavity. The corresponding potential energy contour plot is given in ~\ref{fig:study1}(b). The top family of curves in Fig.~\ref{fig:HeightVsFrequency}(a) is for levitation above the floor within the gap between the stub and the wall. The bottom family of curves is for levitation above the surface of the stub. For example, as shown for the "Above stub" curves in Fig.~\ref{fig:HeightVsFrequency}(a), the cavity's resonant frequency is down-shifted, compared with the empty cavity value, by $\sim50$~MHz when the magnet rests on the end of the stub (i.e., the magnet height is zero). If the magnet is near the edge of the stub, the down-shift will be larger than if the magnet is near the center of the stub. The resonant frequency increases suddenly when the magnet lifts up from the stub. On the other hand, if the magnet is initially placed on the floor of the cavity, within the gap between the stub and the wall, then the resonance frequency is initially up-shifted by $\sim10$~MHz and gradually decreases as the magnet lifts off the floor. The steady-state value of the resonance frequency then depends on the steady-state position of the levitated magnet.  

In our experiments the cavity-magnet system was placed in a dilution refrigerator and we experimentally measured the resonant frequency while the temperature varied between 100~mK and 2~K. We expect, based on our multi-loop-model calculations, that the magnet will levitate within one of two stable wells. Interestingly, this corresponds approximately to the location where the two families of curves in Fig.~\ref{fig:HeightVsFrequency}(a) cross. Experimental measurements of the resonance frequency as a function of temperature are shown in Fig.~\ref{fig:HeightVsFrequency}(b). The magnet was initially placed at the center of the stub. Our model suggests that as the temperature fell, the magnet moved toward the edge of the stub when the temperature was between 700~mK and 800~mK. When the temperature dropped further below 700~mK our model could not differentiate between two possibilities. The magnet either lifted up into the well located at a height of 6.8 mm, or fell into the gap where the well is located at a height of 2.2 mm and remained in that position for the remainder of the experiment. 

The wider-gap (7 mm) cavity has only a single potential well. Figure~\ref{fig:HeightVsFrequency}(c) shows the frequency shift as a function of temperature for a spherical N52 magnet withing the 7-mm radius cavity. Starting from base temperature of about 150 mK, The resonance frequency increases from its low-temperature value by roughly 3~MHz as the magnet drops down from its peak levitation height of about 2.1~mm. Cooling data are given as well indicating a small amount of hysteresis in our measurements. We do not know the origin of the hysteresis, whether it arises from a lag in cooling or some magnet-cavity interaction, but we attribute it here to experimental measurement uncertainty.  

\subsection{Tilting the Cavity}       
\begin{figure}
\includegraphics[scale=.7]{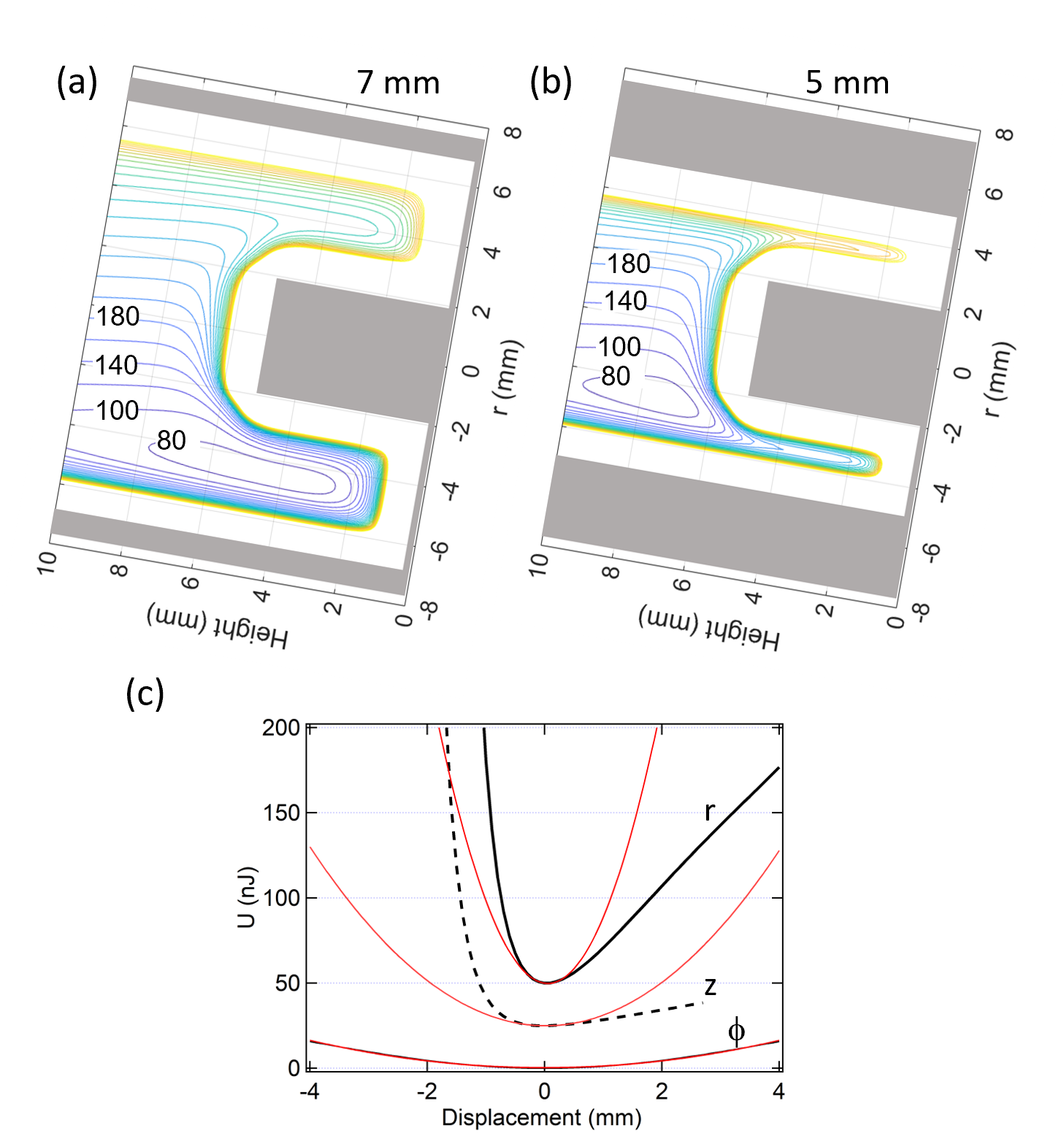}
\caption{\label{fig:TiltedCavities}a) A 7-mm radius cavity tilted up by 10 degrees exhibits a single minimum in the potential energy which is located within the gap. (b) For the case of a narrower, 5-mm cavity, tilted by the same amount, the single minimum moves toward the open end of the cavity. The contour lines in (a) and (b) are spaced by 20 nJ. (c) Plots of the potential energy as a function of displacement from the global minimum equilibrium for the $r$, $z$, and $\phi$ modes. Calculation results are in black and the red curves are quadratic fits to the 3 points at the bottom of the well. These are for the tilted 5-mm cavity and the $r$ and $z$ curves have been offset vertically from one another for clarity.  }
\end{figure}

With the cavity resting perfectly vertically, the magnet may levitate anywhere as a function of angle, $\phi$, around the axis. To configure the experiment to obtain a single and stable equilibrium levitation point, one option is to lay the cavity on its side and tilt it upwards slightly as shown in Fig.~\ref{fig:TiltedCavities}. The resulting potential energy contours are obtained from the simulation data presented in Fig.~\ref{fig:study1} by doing a coordinate rotation when calculating the gravitational part of the potential energy. The diamagnetic response of the cavity to the magnet does not change with tilt. Figure~\ref{fig:TiltedCavities} shows calculations for a N52 magnet within (a) a 7-mm radius cavity, and (b) a 5-mm radius cavity, each tilted upwards as shown by 10 degrees. In both cases we obtain a single minimum in the potential energy, but the minimum is located within the gap for wider cavity (Fig.~\ref{fig:TiltedCavities}(a)) and beyond the stub for the narrower cavity (Fig.~\ref{fig:TiltedCavities}(b)).   

As shown in Fig.~\ref{fig:TiltedCavities}(c), the N-loop model can also be utilized to estimate the resonant mechanical frequency at which the magnet oscillates. When looking at the point in the cavity which yields the lowest potential energy, that position can be combined with the adjacent positions to fit the potential energy versus position (vertical or lateral) using the quadratic function $U(x)=k x^2 + b$ where $x=(r,z,\phi)$ is the displacement from the minimum point, b is an offset added to vertically separate the curves, and k is an effective spring constant. The vibrational modes in the vertical and lateral directions are taken to be independent from one another, the mechanical frequency is given by $f_{mech} = \frac{1}{2\pi}\sqrt{\frac{k}{m}}$. The fits are shown as black lines. For the cylindrical coordinate system of the 5-mm cavity (r, z and $\phi$) containing a 3.67-mg spherical magnet, the estimated resonance frequencies are: $f_r$=28 Hz, $f_z$=11 Hz, and $f_\phi$=4.3 Hz. 

\begin{acknowledgements}
The authors would like to thank Yaniv Rosen and Jonathan DuBois from Lawrence Livermore National Laboratory. Contributions from L. A. Martinez, and A. Castelli were performed under the auspices of the U.S. Department
of Energy by Lawrence Livermore National Laboratory under Contract DE-AC52-07NA27344 (LLNL-JRNL-846269).

\end{acknowledgements}

\appendix

\section{}
\begin{figure}
\includegraphics[scale=0.11]{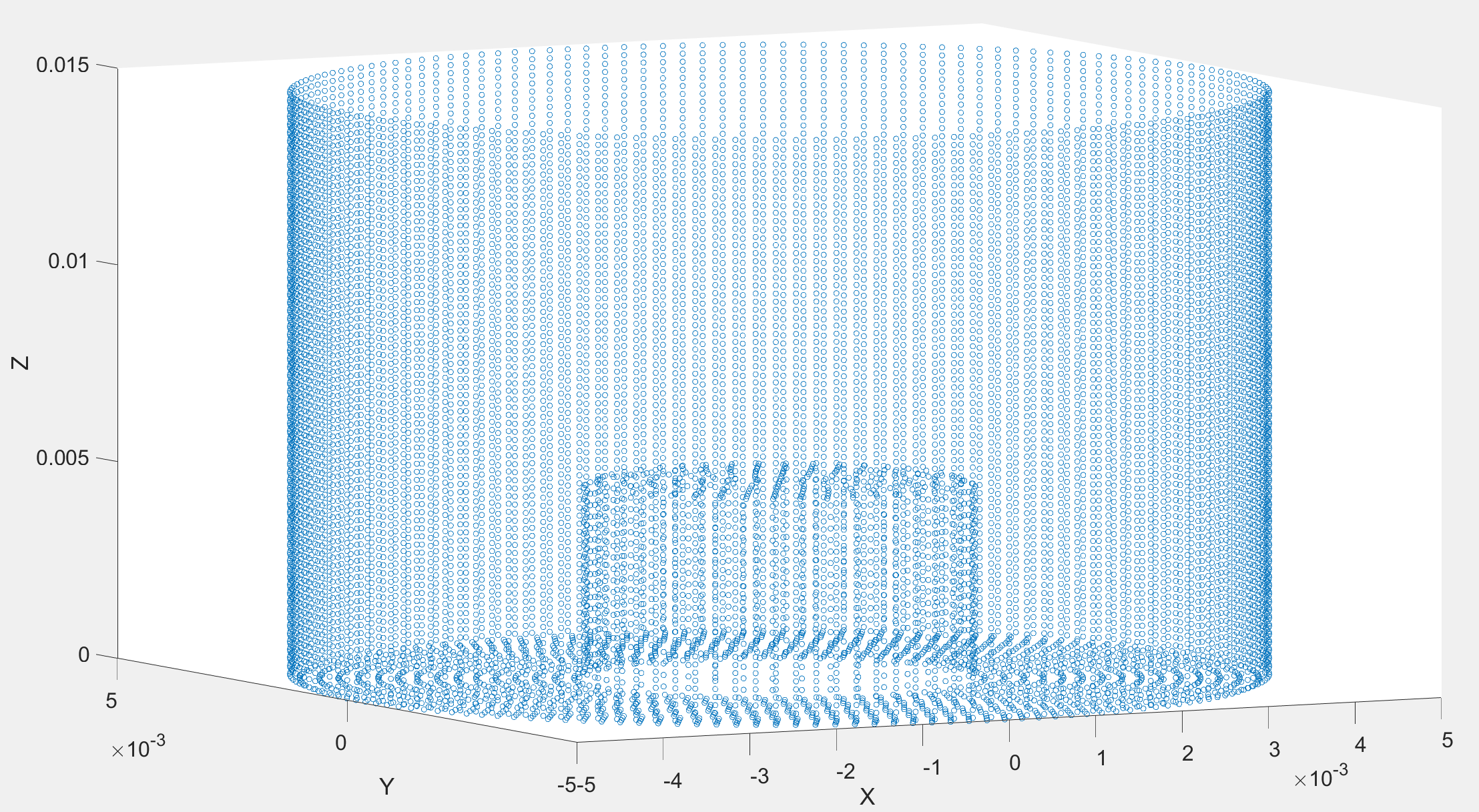}
\caption{\label{fig:Mesh}Visual reference for the mesh generation. Each point represents the center of a circular mesh element used in the calculation. The four surfaces (stub top, cavity floor, stub wall, cavity wall) are generated separately and all surface elements have roughly the same radius between the four surfaces. The resolution (spacing between the center point of each circular mesh element) is easily changed for more detailed calculations.} 
\end{figure}

Two separate programs were written to accomplish this task. First, the cavity meshing needs to be generated. To do this, the four surfaces of the coaxial cavity (coaxial stub top, coaxial stub curved surface, cavity floor, outer radius curved surface) are broken up and meshed individually. For the flat surfaces, a script that fills a large circle with equally sized smaller circles is used to generate approximately equal surface elements. The center coordinates of each loop and the corresponding radius are stored, along with the surface normal vector in the frame of reference of the cavity (lab frame). The mesh generation is general and can be adapted to work with other cavity geometries. The resolution can also be changed to allow for finer or coarser simulations. An example of the meshing is shown in Fig.~\ref{fig:Mesh}.

\begin{figure}[ht]
\includegraphics[scale=1]{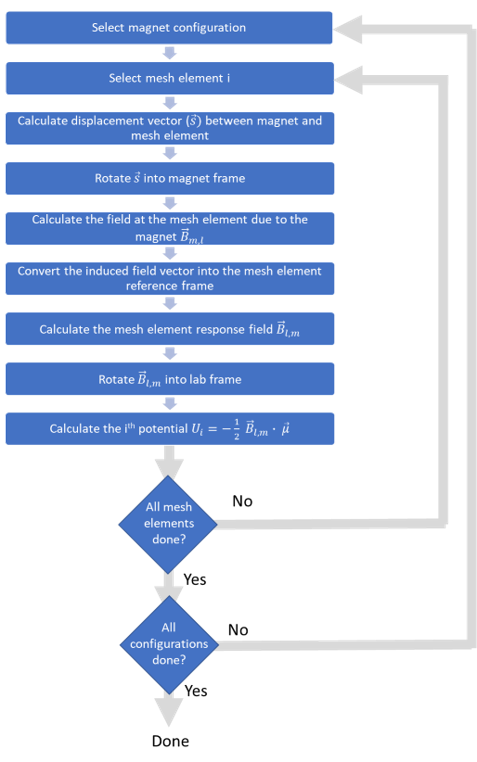}
\caption{\label{fig:flowchart}a) Flowchart describing the algorithm used to calculate the magnetic potential energy of the magnet. Each rotation around the chart represents the magnetic potential energy of the magnet due to a single loop. This is repeated for all loops and added together to get the total magnetic potential energy for each magnet position.}
\end{figure}

Once the mesh is generated and the necessary quantities are known (coordinates, loop radius, and surface normal vector) for each mesh element, the second program is used to loop through each individual element and calculate the magnetic potential energy for that particular magnet position. The algorithm for this involves several changes of reference frame and a number of vector rotations. This algorithm is described visually in Fig.~\ref{fig:flowchart}. Because of the three different reference frames (cavity, magnet, mesh element) involved in this calculation, great care had to be taken when rotating vectors between coordinate systems.
    
Once all of the individual mesh element potentials are calculated for a given magnet position, each contribution is summed for the total potential energy over the whole cavity. With $N$ loop elements, the total potential energy $U$ is given by
    
    \begin{equation}
        U_{total} = \left(\sum_{i=1}^N U_i\right) + mgh
    \end{equation}

The program that loops through each mesh element will repeat this calculation for a range of magnet orientations (spatial position and angular tilt from the z-axis of the cavity) and save the total potential energy for each orientation. Doing so allows for the generation of a contour plot showing the total potential energy of the magnet-cavity system for a range of magnet positions and orientations. The point of minimum potential is then calculated. Figure~\ref{fig:flowchart} shows a flowchart of the algorithm used to find the total potential energy for each magnet position. Figures~\ref{fig:study1}(a) and (b), for example, show typical contour plots describing the potential energy landscape of an N52 ($B_r = 1.47 T$) magnet with a moment aligned with the positive z-axis generated by the N-loop model.

\nocite{*}
*\bibliography{potentialenergybib}
\end{document}